\input{aipcheck}

\documentclass[
    ,final            
  ]
  {aipproc}

\layoutstyle{6x9}

\begin{document}

\title{Gluon Dominance Model}

\classification{ \texttt{13.65+i, 13.75.-n, 13.85.-t, 13.85.Rm}}

\author{Kokoulina E.S.}{
  address={JINR, Dubna, Moscow region, \\
  141980, Russia \\
  GSTU, LPS,
  October av. 48, Gomel \\
  246746, Belarus \\
  } }

%

\begin{abstract}
A new way to study the multiplicity production in high energy
processes is proposed. It is based on the multiplicity
distribution description by the schemes taking into account the
formation of quark-gluon system and hadronization. This
investigations revealed an active role of gluons and the universal
mechanism of their hadronization.
%
%
\end{abstract}

\maketitle


\section{Introduction}

At present there is neither single consistent theory nor
convincing model which could explain all the results obtained at
RHIC and SPS \cite{WhPa}. It is evident that in these experiments
the local thermal equilibrium is reached and thermalized matter is
produced. Thermal statistical models based on this approach
explain yields of different hadrons at high energy \cite{WhPa}. At
the same time these models describe well the particle yields in
$e^+e^-$ and $pp$ interactions where, as generally accepted,
thermalization is impossible \cite{Khar}. So the intensive study
and insight of these processes remain to be very importance.

To investigate multiparticle production (MP) at high energy in
these processes, a two stage model was proposed \cite{TSM,MGD}. It
is based on the use of QCD and the phenomenological scheme of
hadronization. The model describes well multiplicity distributions
(MD) and their moments in $e^+e^-$ annihilation, $pp$ and $p\bar
p$ interactions and gives complementary information to a better
understanding of MP of relativistic heavy ion collisions.

This model confirms the fragmentation mechanism of hadronization
in $e^+e^-$ annihilation and the transition to recombination
mechanism in hadron and nucleus interactions. It explains the
shoulder structure in MD at higher energies and the behavior of
$f_2$ in $p \bar p$ annihilation at few tens GeV/c by including
intermediate quark topologies. The mechanism of the soft photons
(SP) production as a sign of hadronization and estimates of their
emission region size is proposed \cite{MGD1}.

The $e^+e^-$ annihilation is the most suitable process to study
MP. It can be realized through the formation of virtual $\gamma$
or $Z^0$--boson which then decays into two quarks:
$e^+e^-\rightarrow(Z^0/\gamma)\rightarrow q\bar q$. The
$e^+e^-$--reaction is simple for analysis, as the produced state
is pure $q\overline q$. To describe the process of parton fission
(quarks and gluons) at big virtuality, pQCD may be applied. This
stage can be named as the stage of cascade. When partons get small
virtuality, pQCD cannot be applied. Therefore phenomenological
models are used to describe hadronization in this case.

The probabilistic nature of parton fission in QCD has been
established in \cite{KUV}.

\section{$e^+e^-$ - annihilation}

A.Giovannini~\cite{GIO} had proposed to describe quark and gluon
jets as markovian branching processes with three elementary
contributions: gluon fission, quark bremsstrahlung and quark pair
production. He constructed a system of differential equations for
generating functions (GF) and obtained solutions of MD for quark
jet
\begin{equation}
\label{4} P_{m}^{(q)}(Y)=\frac{k_p(k_p+1)\dots(k_p+
m-1)}{m!}\left(\frac{\overline m} {\overline
m+k_p}\right)^{m}\left( \frac{k_p}{k_p+\overline m}\right)^{k_p}.
\end{equation}
These MD are known as negative binomial distribution (NBD). The GF
for them is $Q^{(q)}(z,Y)=\sum\limits_{m=0}^{\infty} z^{m}
P_{m}(Y)= \left[1+ \overline m/k_p (1-z)\right]^{-k_p}$, where
$\overline m$ is the mean gluon multiplicity, $Y-$QCD evolution
variable and $k_p$ - NBD parameter.

In \cite{TSM} MD (\ref{4}) was taken for the cascade stage. For
the hadronization a sub narrow binomial distribution (BD) was
added

\begin{equation}
\label{12} P_n^H=C^n_{N_p}\left(\frac{\overline n^h_p}
{N_p}\right)^n\left(1-\frac{\overline n_p^h}
{N_p}\right)^{N_p-n},\quad Q^H_p(z)=\left[1+\frac{\overline n^h_p}
{N_p}(z-1)\right]^{N_p},
\end{equation}
where $C_{N_p}^n$ - binomial coefficient, $\overline n^h_p$ and
$N_p$ ($p=q,g$) have a sense of mean multiplicity and maximum
number of secondary hadrons formed from parton (p) at its passing
of the second stage.

We have chosen BD from the analysis of experimental data on
$e^+e^-$- annihilation. Second correlation moments $f_2$ were
negative at low energies (less than 9 GeV). The choice of such
distributions was the only one that could describe the experiment.
We suppose  that the hypothesis of soft decoloration is right.
Therefore the hadronization stage is added to the parton stage by
means of a factorization. We introduce parameter $\alpha=N_g/N_q$
to distinguish the hadrons, produced from quark or gluon. MD of
hadrons in $e^+e^-$ - annihilation can be written as follows
($N_q=N$, $\overline n^h=\overline n_q^h$):
\begin{equation}
\label{15} P_n(s)= \sum\limits_{m=0} ^{M_g}P_m^{(q)}C^n_{(2+\alpha
m)N} \left(\frac{\overline n^h} {N}\right)^n\left(1-\frac
{\overline n^h}{N}\right)^{(2 +\alpha m)N-n}.
\end{equation}

The expression (\ref{15}) describes well the experimental data
\cite{DAT} from 14 to 189 GeV \cite{NPCS} (e.g. Fig. 1). The mean
gluon multiplicity $\overline m$ has a tendency to rise, but
slower than the logarithmic one. Parameter $k_p$ was been related
with temperature in \cite{TEM}. The fact that $\alpha <1$ is the
evidence that hadronization of a gluon is softer than a quark. It
is surprising that gluon parameters of hadronization ($N_g$,
$\overline n^h_g$) remain constant without considerable deviations
in spite of the indirect finding: $N_g\sim 3-4$ and $\overline
n^h_g\sim 1$ (e.g. Fig. 2). Therefore we can draw a conclusion
about the universality of gluon hadronization in $e^+e^-$
annihilation in the sufficient wide energy region.

It was shown \cite{OSC} that the ratio of factorial cumulative
moments over factorial moments changes the sign as a function of
the order. The calculation of this ratio by using (\ref{15}) was
done in \cite{TSM}. It has been obtained that the period of
oscillations is equal to 2 before $Z^0$-region and increases at
higher energies.

\begin{ltxfigure}
\begin{minipage}[b]{.3\linewidth}
\centering
\includegraphics[width=\linewidth, height=2in, angle=0]{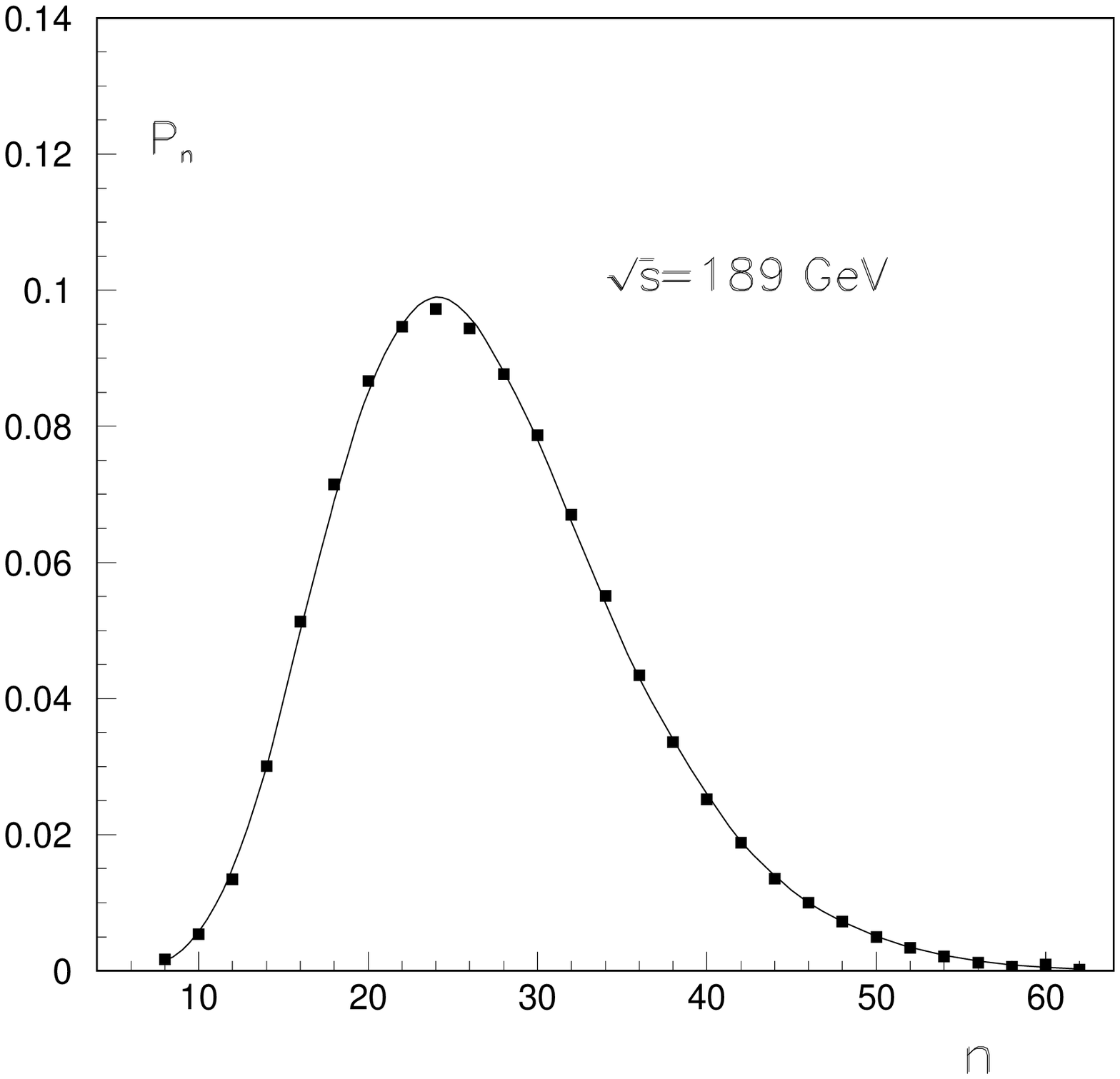}
\caption{MD at 189GeV.} \label{13dfig}
\end{minipage}\hfill
\begin{minipage}[b]{.3\linewidth}
\centering
\includegraphics[width=\linewidth, height=2in, angle=0]{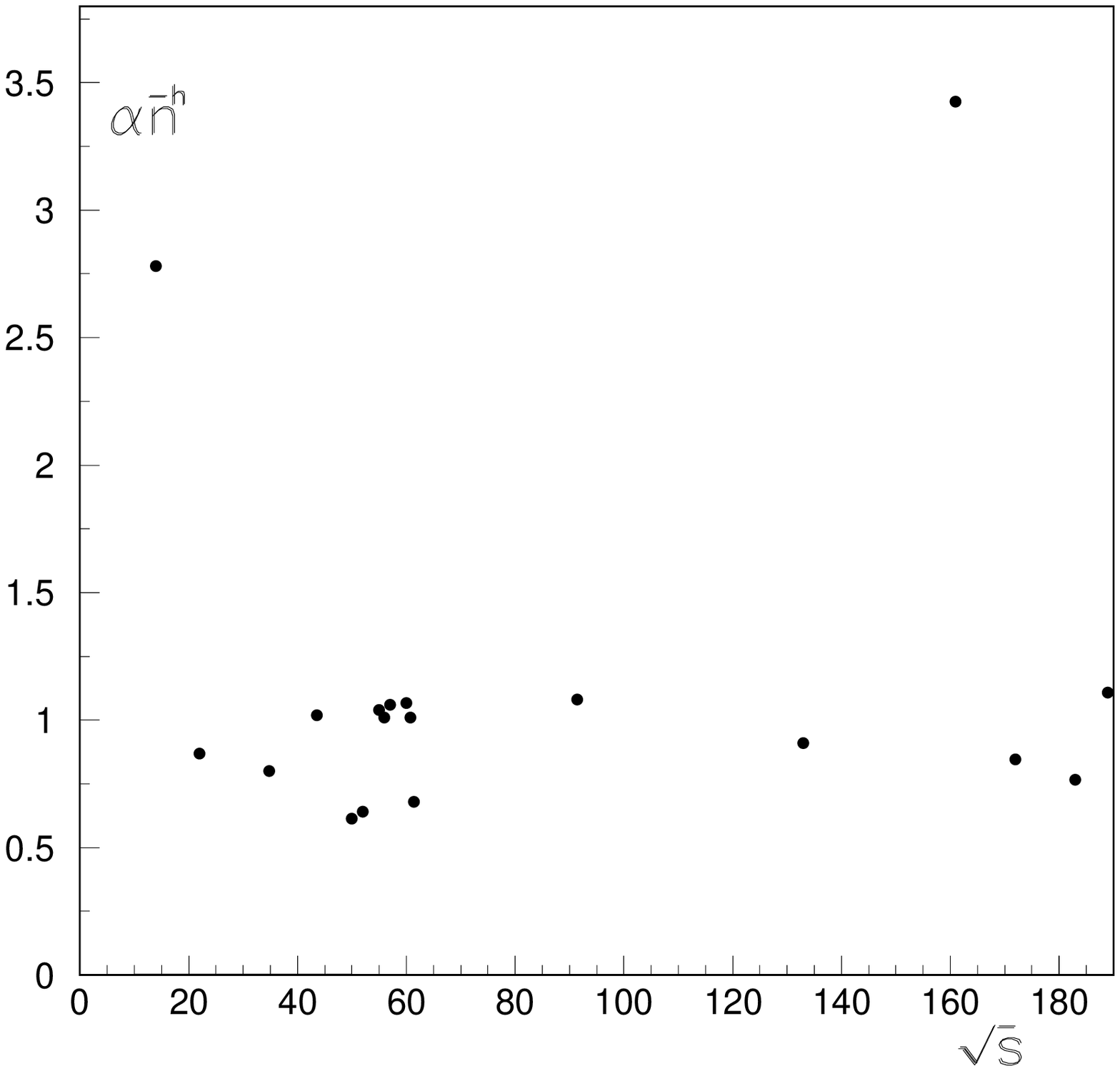}
\caption{$\overline n_g^h=\alpha \overline n_q^h$ .}
\label{14dfig}
\end{minipage}\hfill
\begin{minipage}[b]{.3\linewidth}
\centering
\includegraphics[width=\linewidth, height=2in, angle=0]{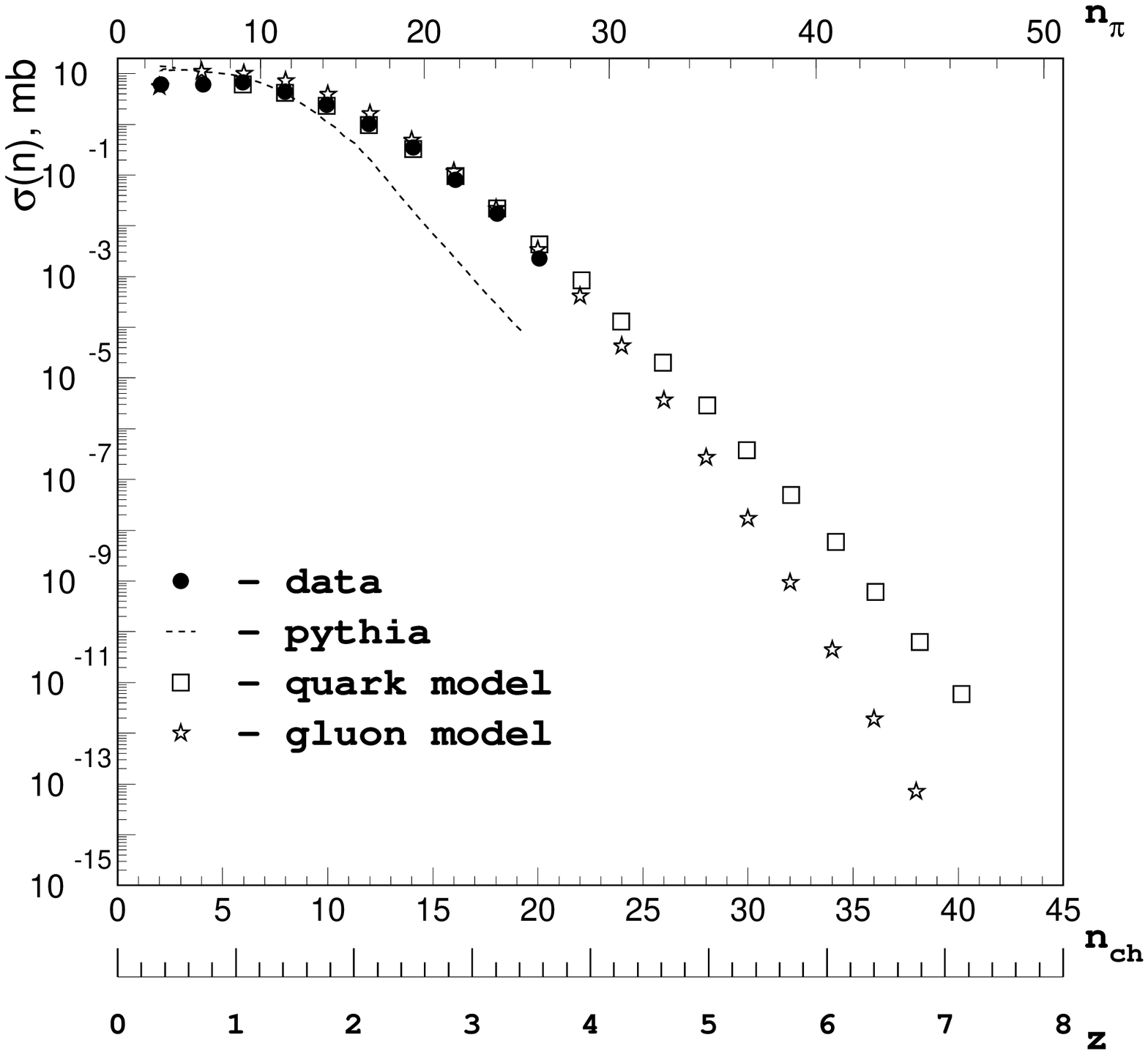}
\caption{$\sigma (n)$ in $pp$.} \label{15dfig}
\end{minipage}
\end{ltxfigure}

\begin{ltxfigure}
\begin{minipage}[b]{.3\linewidth}
\centering
\includegraphics[width=\linewidth, height=2in, angle=0]{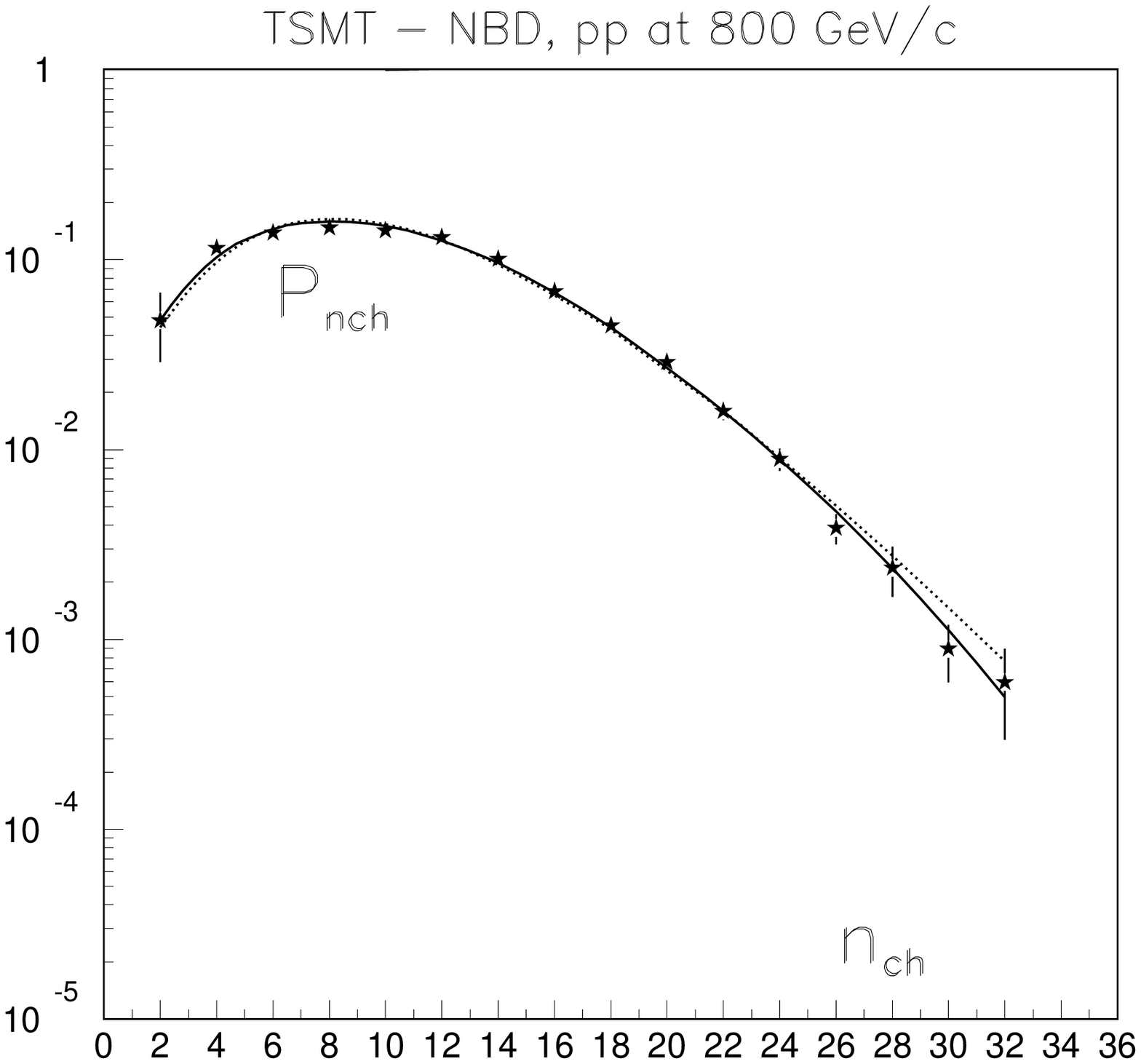}
\caption{MD in GDM and NBD.} \label{13dfig}
\end{minipage}\hfill
\begin{minipage}[b]{.3\linewidth}
\centering
\includegraphics[width=\linewidth, height=2in, angle=0]{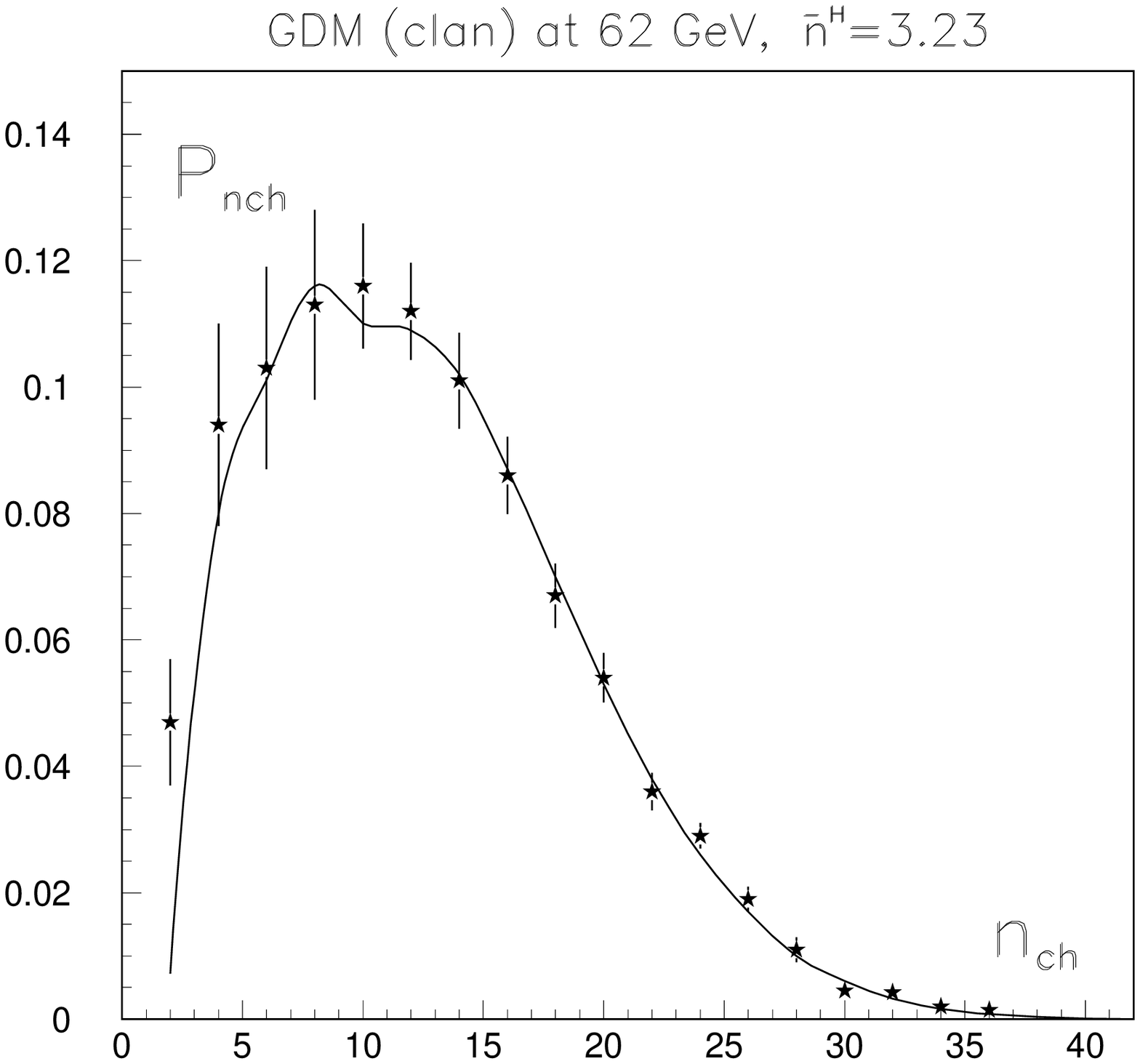}
\caption{MD in GDM (clan).} \label{14dfig}
\end{minipage}\hfill
\begin{minipage}[b]{.3\linewidth}
\centering
\includegraphics[width=\linewidth, height=2in, angle=0]{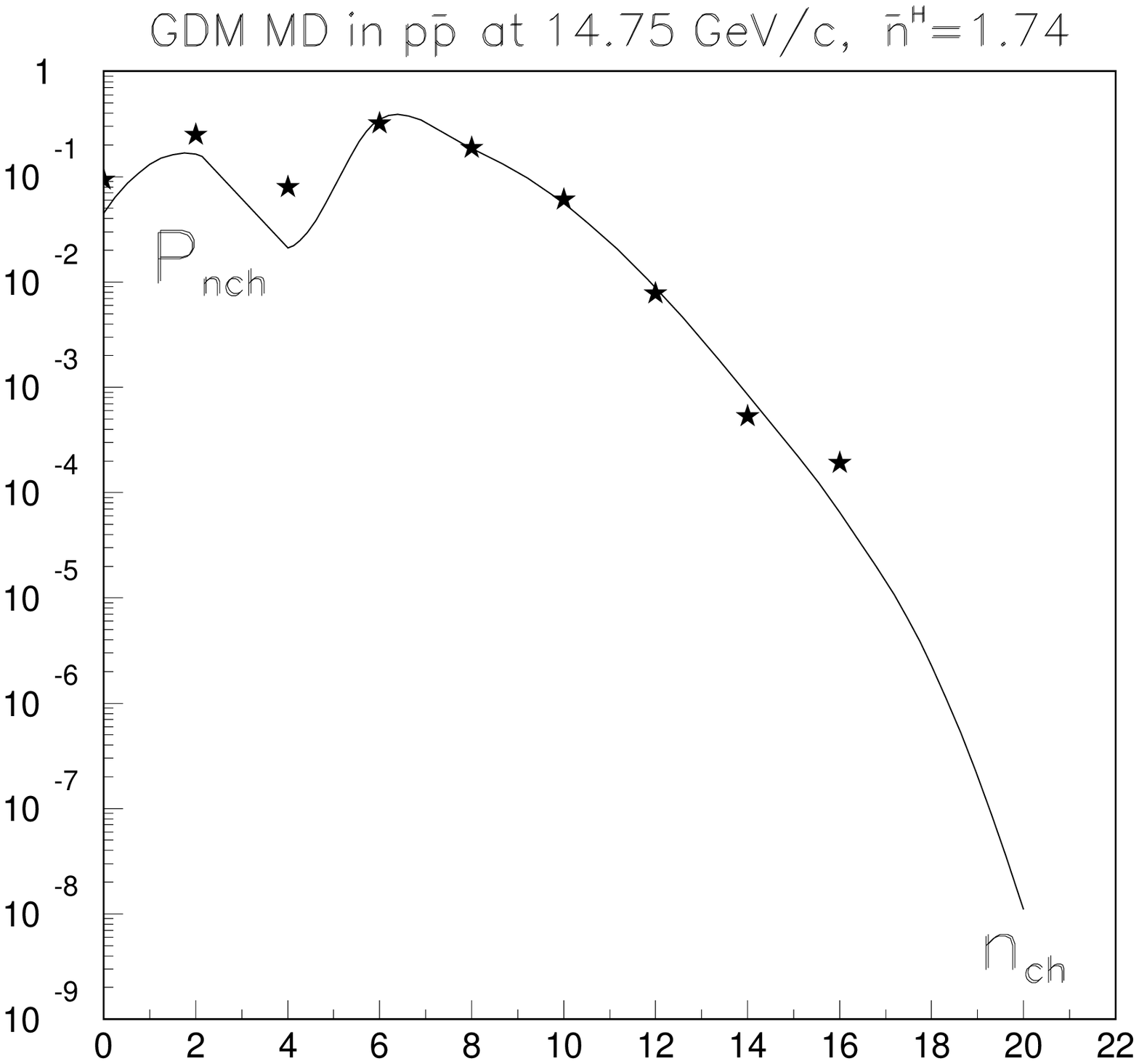}
\caption{MD in $p\overline p$ at 14.75 GeV/c and in GDM .}
\label{15dfig}
\end{minipage}
\end{ltxfigure}
\section{$pp$ interactions}
The study of MP in $pp$ interactions is implemented in the
framework of the project "Thermalization" \cite{THE}. This project
is aimed at studying the collective behavior of secondary
particles and advancement to the high multiplicity region (HMR)
beyond available data \cite{BAB} in $pp$ interactions at 70 GeV/c.
The calculation by the MC PHYTHIA code has shown that the standard
generator predicts a value of the cross section which is in a
reasonably good agreement with data at small multiplicity
$n_{ch}<10$, but it underestimates the value $\sigma(n_{ch})$ by
two orders of the magnitude at $n_{ch}=20$. The existing models
are very much sensitive in this region \cite{CHI}, also (Fig. 3).

We suppose that after the inelastic collision the part of the
energy of the initial impact protons is transformed to the inside
energy. Several quarks and gluons become free and form quark-gluon
system (QGS). Partons which can produce hadrons are named the
active ones. Two schemes were proposed \cite{MGD}. In the first
scheme the parton fission inside the QGS is taken into account
(the scheme with a branch). If we are not interested in what is
going inside QGS, we come to the scheme without a branch (TSMT).

At the beginning of research we took a model with active quarks
and gluons. Parameters of that model had values which differed
very much from parameters obtained in $e^+e^-$- annihilation,
especially for hadronization. It was one of the main reasons to
refuse the scheme with active quarks. So, we reserve quarks
remained inside of the leading particles. All of the newly born
hadrons were formed by active gluons. That is why we began to name
this model -- the gluon dominance model (GDM).

The Poisson distribution was chosen as the simplest MD for active
gluons which appeared for the first time after the collision. The
number of these gluons fulfils the role of the impact parameter
for nucleus. To describe MD in the gluon cascade (fission), Farry
distribution was used with GF $Q_k^B(z)= z^k/\overline
m^k\left[1-z \left(1-\frac{1}{\overline m} \right)\right]^{-k}$,
where $k$ -- the number of initial gluons in QGS, $\overline m$ --
the mean multiplicity of them in the end of the branch. On the
second stage some of active gluons can leave QGS ("evaporate") and
transform to real hadrons. Our BD (\ref{12}) for hadrons on the
stage of hadronization is as follows
\begin{equation}
\label{34} P_n^H(m)= C^{n-2}_{\delta  mN_g}\left(\frac{\overline
n^h_g} {N_g}\right)^{n-2}\left(1-\frac{\overline n^h_g}
{N_g}\right)^{\delta mN_g-(n-2)}.
\end{equation}
In (\ref{34}) parameter $\delta $ is the ratio of evaporated
gluons leaving QGS, to all active gluons. From the comparison with
data \cite{BAB} we have obtained that a maximum possible number of
hadrons from a single gluon looks very much like the number of
partons in the glob of cold QGP of L.Van Hove \cite{LVH}, the
branch processes are weak. The fraction of released gluons is
equal to $\delta = 0.47 \pm 0.01$ (the same as in \cite{AHM}). A
part of active gluons does not convert into hadrons. They stay in
QGS and can become sources of soft photons (SP).

In the scheme without a branch we consider that evaporated gluons
have Poisson MD, too. Using the idea of the convolution of two
stages $P_n(s)=\sum \limits _{m} P_m^P(s)P_n^H(m)$ as well as the
BD for hadronization, we obtain MD in pp-collisions. The
comparison GDM with the data \cite{BAB} (Fig. 3) gives values of
parameters $N_g=4.24 \pm 0.13$, $\overline n^h_g=1.63\pm 0.12$ and
it is in agreement with the values obtained in $e^+e^-$
annihilation \cite{NPCS}. We are restricted in sum to $m=6$. Our
estimation of the maximal possible observable number of charged
particles is $\leq 26$.

We have also got MD for neutral mesons and total multiplicity by
using mean multiplicities of $\pi ^0$-mesons \cite{MUR} and active
gluons, expected approximate equality of probabilities of the
formation charged and neutral particles from single gluon at
hadronization and the above-mentioned idea of the convolution
\cite{MGD1}. The maximum observable number of them is estimated as
16, total - as 42, the parameter $\overline n^h_g=1.036\pm 0.041$
(h = $\pi ^0$).

The analysis of the mean multiplicity of $\pi^0$-- mesons  versus
the number of charged particles $n_{ch}$, allows to set
limitations to the number of neutral mesons at given $n_{ch}$ and
indicates the absence of AntiCentauro events (Centauro may be in
HMR). The obtained estimations of probabilities for charged and
neutral hadrons production from gluon while its passing through
hadronization permits to get "the charged hadron/pion" ratio
\cite{PHE} in pp interactions. At 69 GeV/c this ratio is equal to
$1.19\pm 0.25$.

GDM describes well MD in the region of $100-800$ GeV/c (Fig. 4 and
table 1 in \cite{MGD1}). The number of active gluons, their mean
multiplicity, $N_g$ and $\overline n^h_g$ increase slowly. A
growth of $\overline n^h_g$ in $pp$ interactions indicates a
possible change mechanism of hadronization of gluons in comparison
with $e^+e^-$ annihilation. It is considered that in the last case
the partons transform to hadrons by the fragmentation mechanism at
the absence of the thermal medium. Our MD analysis gives
$\overline n_g^h \sim 1$ for this hadronization \cite{NPCS}. The
recombination is specific for the hadron and nucleus processes. In
this situation a lot of quark pairs from gluons appear almost
simultaneously and recombine to various hadrons \cite{HWA}. The
value $\overline n_g^h$ becomes bigger $\sim 2-3$, that indicates
this transition. The recombination mechanism provides
justification for applying the statistical model to describe
ratios of hadron yields and the explanation of the collective flow
of quarks \cite{HWA}.

In \cite{GIO} MP is described by means of a clan mechanism and
emphasizes the gluon nature of clan. Our GDM allows to give a
concrete content for the clan. The clan model uses the logarithmic
distribution (LD) in a single clan. LD are similar to our BD.

At the SPS energy the shoulder structure appears in MD \cite{UA5}.
As it was marked in the branch scheme, the gluon fission is
strengthened at higher energies. The independent evaporation of
gluon sources of hadrons may be realized as single gluons as
groups from two and more fission gluons. Following \cite{GIO} we
name such groups - clans. GDM with two kinds of clans \cite{MGD1}
describes data \cite{ISR} very well (Fig. 4). Moreover "the
charged hadron/pion" ratio at 62 GeV is equal to $\sim 1.6$ and
agrees with Au-Au peripheral interactions at 200 GeV and with pp
interactions at 53 GeV \cite{PHE}.

The specific feature of GDM is the dominance of active gluons in
MP. We expect the emergence of many of them in nucleus collisions
at RHIC and the formation of a new kind of matter (QGP) at high
energy. Our QGS can be a candidate for this.

Experiments \cite{CHL} have shown that the measured cross sections
of SP are several times larger than the expected ones from QED.
The phenomenological glob model explains the SP excess \cite{LVH}.
We consider that at a certain moment the QGS or excited new
hadrons may set in an almost equilibrium state during a short
period of time. That is why, to describe massless photons, we have
used the black body emission spectrum \cite{HER}. At 70 GeV/c an
inelastic cross section is equal to $\sim 40 mb$, the SP formation
cross section is about $4 mb$ \cite{THE} and since $ \sigma
_{\gamma} \simeq n_{\gamma }(T)\cdot \sigma _{in}$ then $n_{\gamma
}\approx 0.1$. If $n_{\gamma }$ and temperature $T(p)$
($p$--momentum) are known, then we can estimate the emission
region size $L$ of SP. The obtained values $L\sim 4-6 fm$
\cite{MGD1, KUR}.

In conclusion one can show how GDM may explain experimental $p\bar
p$ annihilation data at tens GeV/c \cite{Rush}. The differences
between $p\bar p$ and $pp$ inelastic topological cross sections
($\Delta \sigma _n(p\overline p -pp)= \sigma _n(p \overline p) -
\sigma _n (pp))$ point out the contribution of different
mechanisms in MP. The negative values of $f_2$ indicate the
predominance of the hadronization stage in MP. According to MGD,
the active gluons are a basic source of secondary hadrons.

There are three valent $q\overline q$-pairs at the initial stage
of annihilation . They can turn to the "leading" mesons which
consist of: a) valent quarks or b) valent and sea quarks
\cite{KUR}. In the case a) three "leading" neutral pions (the
"0"--topology) or two charged and one neutral "leading" mesons
("2"--topology) may form. In b) case the "4"- and "6"- topologies
are realized. The neutron and antineutron formation is possible,
too.

The active gluons emerge together with the formation of
intermediate topology. At this region hadronization dominates
since $f_2$ is negative. In the simple case for m active gluons GF
$Q_m(z)=[1+\overline n^h/N(z-1)]^{mN},$ and $f_2=-m(\overline
n^h)^2/N$. If $m$ grows while increasing the energy of the
colliding particles, then $f_2$ will decrease almost linearly from
$m$ that agrees with the data \cite{Rush}.

According to GDM \cite{MGD1} and taking into account an
intermediate charged topology ("0", "2" and "4") with  active
gluons, GF $Q(z)$ for final annihilation MD ($\Delta \sigma
_n(p\overline p -pp)/\sum _n \Delta \sigma _n(p\overline p -pp)$)
may be written as the convolution gluon and hadron components
 $$Q(z)=c_0\sum\limits_m^{M_0} P_m^G
[Q^H(z)]^m+c_2z^2\sum\limits_m^{M_2}P_m^G[Q^H(z)]^m
+c_4z^4\sum\limits_m^{M_4}P_m^G[Q^H(z)]^m.$$ The parameters of
$c_0$, $c_2$  and $c_4$ are parts of intermediate topologies. GDM
describes data (Fig. 6) with the ratio $c_0$ : $c_2$ : $c_4$ = $15
: 40 : 0.05$ and $M_0 \sim M_2 \sim 1-2, \quad M_4 \sim
4$~\cite{MGD1}. The carried out investigations allow one to
understand deeper the MP nature.

I thank the Organizing committee for a wonderful possibility to
take part in ISMD2005.



\bibliographystyle{aipproc}   






\end{document}